\documentstyle[referee]{mn}
\input psfig.tex
\def\lsim{\lower.5ex\hbox{$\; \buildrel < \over \sim \;$}}
\def\gsim{\lower.5ex\hbox{$\; \buildrel > \over \sim \;$}}
\def \simeq{\lower.3ex\hbox{$\; \buildrel \sim \over - \;$}}
\def\ch{\lower-0.55ex\hbox{--}\kern-0.55em{\lower0.15ex\hbox{$h$}}}
\def\lh{\lower-0.55ex\hbox{--}\kern-0.55em{\lower0.15ex\hbox{$\lambda$}}}
\def\eg{{\it e.g.,} }
\def\etal{{\em et al.} }
\def\ie{{\em i.e.,} }

\newif\ifAMStwofonts
\ifoldfss
  \ifCUPmtlplainloaded \else
    \NewTextAlphabet{textbfit} {cmbxti10} {}
    \NewTextAlphabet{textbfss} {cmssbx10} {}
    \NewMathAlphabet{mathbfit} {cmbxti10} {} % for math mode
    \NewMathAlphabet{mathbfss} {cmssbx10} {} %  "   "    "
  \fi
  \ifAMStwofonts
    \ifCUPmtlplainloaded \else
      \NewSymbolFont{upmath} {eurm10}
      \NewSymbolFont{AMSa} {msam10}
      \NewMathSymbol{\upi}     {0}{upmath}{19}
      \NewMathSymbol{\umu}     {0}{upmath}{16}
      \NewMathSymbol{\upartial}{0}{upmath}{40}
      \NewMathSymbol{\leqslant}{3}{AMSa}{36}
      \NewMathSymbol{\geqslant}{3}{AMSa}{3E}

      \let\leq=\leqslant 
       
    \fi
  \fi
\fi % End of OFSS
\ifnfssone
  \newmathalphabet{\mathit}
  \addtoversion{normal}{\mathit}{cmr}{m}{it}
  \addtoversion{bold}{\mathit}{cmr}{bx}{it}
  \newmathalphabet{\mathbfit} % math mode version of \textbfit{..}
  \addtoversion{normal}{\mathbfit}{cmr}{bx}{it}
  \addtoversion{bold}{\mathbfit}{cmr}{bx}{it}
  \newmathalphabet{\mathbfss} % math mode version of \textbfss{..}
  \addtoversion{normal}{\mathbfss}{cmss}{bx}{n}
  \addtoversion{bold}{\mathbfss}{cmss}{bx}{n}
  \ifAMStwofonts
    \ifCUPmtlplainloaded \else
      \UseAMStwoboldmath
      \makeatletter
      \new@mathgroup\upmath@group
      \define@mathgroup\mv@normal\upmath@group{eur}{m}{n}
      \define@mathgroup\mv@bold\upmath@group{eur}{b}{n}
      \edef\UPM{\hexnumber\upmath@group}
      \new@mathgroup\amsa@group
      \define@mathgroup\mv@normal\amsa@group{msa}{m}{n}
      \define@mathgroup\mv@bold\amsa@group{msa}{m}{n}
      \edef\AMSa{\hexnumber\amsa@group}
      \makeatother
      \mathchardef\upi="0\UPM19
      \mathchardef\umu="0\UPM16
      \mathchardef\upartial="0\UPM40
      \mathchardef\leqslant="3\AMSa36
      \mathchardef\geqslant="3\AMSa3E

      \let\leq=\leqslant 

    \fi
  \fi
\fi % End of NFSS release 1

\ifnfsstwo
  \DeclareMathAlphabet{\mathbfit}{OT1}{cmr}{bx}{it}
  \SetMathAlphabet\mathbfit{bold}{OT1}{cmr}{bx}{it}
  \DeclareMathAlphabet{\mathbfss}{OT1}{cmss}{bx}{n}
  \SetMathAlphabet\mathbfss{bold}{OT1}{cmss}{bx}{n}
  \ifAMStwofonts
    \ifCUPmtlplainloaded \else
      \DeclareSymbolFont{UPM}{U}{eur}{m}{n}
      \SetSymbolFont{UPM}{bold}{U}{eur}{b}{n}
      \DeclareSymbolFont{AMSa}{U}{msa}{m}{n}
      \DeclareMathSymbol{\upi}{0}{UPM}{"19}
      \DeclareMathSymbol{\umu}{0}{UPM}{"16}
      \DeclareMathSymbol{\upartial}{0}{UPM}{"40}
      \DeclareMathSymbol{\leqslant}{3}{AMSa}{"36}
      \DeclareMathSymbol{\geqslant}{3}{AMSa}{"3E}

      \let\leq=\leqslant 

    \fi
  \fi
\fi % End of NFSS release 2

\ifCUPmtlplainloaded \else
  \ifAMStwofonts \else % If no AMS fonts
    \def\upi{\pi}
    \def\umu{\mu}
    \def\upartial{\partial}
  \fi
\fi

\title{Radiatively driven electron-positron jets from two component
accretion flows.}
\author[Chattopadhyay \etal]
       {Indranil Chattopadhyay$^1$, Santabrata Das$^2$ and Sandip K. Chakrabarti$^{2,3}$\\
$^1$ Centre for Plasma Astrophysics, Department of Mathematics, K. U. Leuven,\\
Celestijnenlaan 200B, Leuven 3001, Belgium;\\
$^2$ S. N. Bose National Centre for Basic Sciences, JD Block, Sector-III,
SaltLake, Kolkata 700098, India; \\
$^3$ Also at Centre for Space Physics, P61, Southend Garden, Calcutta 700084, India\\
Indranil.Chattopadhyay@wis.kuleuven.ac.be, sbdas@bose.res.in, chakraba@bose.res.in}

\date{Accepted .
      Received ;
      in original form }
\pubyear{}
\begin{document}

\maketitle

\begin{abstract}

Matter accreting onto black holes has long been known to have standing 
or oscillating shock waves. The post-shock matter puffs up in the form
of a torus,
which intercepts soft photons from the outer Keplerian disc and inverse
Comptonizes
to produce hard photons. The post-shock region also produces jets.
We study the interaction of both hard photons and soft photons,
with on-axis electron-positron jets. We show that the radiation from
post-shock torus accelerates the flow to relativistic velocities,
while that from the Keplerian disc has marginal effect. We also show that,
the velocity at
infinity or terminal velocity ${\vartheta}$, depends on the shock location
in the disc. 
\end{abstract}


\section{Introduction}

Jets around quasars and micro-quasars show relativistic terminal speeds.
While jets are quite ubiquitous and are associated with a wide range of
celestial objects, only some jets around quasars and micro-quasars
show highly relativistic terminal speed (\eg GRS 1915+105, Mirabel
{\&} Rodriguez 1994; 3C 273, 3C 345, Zensus \etal 1995; M87, Biretta 1993). 
These relativistic jets are generally associated with compact objects
and circumstantial evidences show that many of these central 
gravitating objects are black holes. Black holes do not have 
`hard surfaces' nor do they have atmospheres, hence 
if observations show that many of these jets come from the vicinity
of the black hole, then they must originate from the accretion discs around
these black holes.

Inner boundary conditions of matter accreting onto a black hole are (i)
supersonic and 
(ii) sub-Keplerian.  Liang and Thompson (1980) showed that sub-Keplerian
matter accreting onto a black hole has at least two X-type critical points.
In much of the parameter space, it has been shown that the
supersonic matter crossing the outer critical point, under
goes centrifugal pressure mediated shock (Fukue, 1987;
Chakrabarti, 1989),
becomes subsonic and enters the black hole through the inner X-type
critical point.
Entropy is generated at the shock making the post shock region hot.  
The region in which the flow slows down may be extended if the shock 
conditions are not satisfied. This hot, slowed down region is puffed up in
the form of a torus (hereafter, CENBOL ${\equiv}$
CENtrifugal pressure supported BOundary Layer; see Chakrabarti \etal 1996,
hereafter CTKE96).
This disc due to its advection term may be called advective accretion
disc (Chakrabarti 1989, hereafter C89; Chakrabarti 1990,
hereafter C90; Chakrabarti 1996, hereafter C96;
CTKE96).
Chakrabarti and Titarchuk (1995, hereafter CT95),
proposed a disc model which contains both the Keplerian matter and the
sub-Keplerian 
matter. In this model, the Keplerian matter is of higher angular momentum
and low
specific energy, and settles around the equatorial plane to form the
Keplerian disc (see, Shakura \& Sunyaev 1973; Novikov
\& Thorne 1973, hereafter NT73) while the sub-Keplerian matter with high energy
and lower viscosity flanks the cooler Keplerian disc from the top and bottom,
sandwiching the Keplerian disc known as the sub-Keplerian halo (see,
CT95; Chakrabarti 1997, hereafter C97)
in the literature. The sub-Keplerian halo may suffer a standing shock at
$x_s$, a few tens of 
Schwarzschild radii,  and it may be sustained there if the post-shock
thermal pressure is
sufficiently high. The shock compresses the pre-shock flow making it denser and 
at the same time hotter. In the model solution proposed by Chakrabarti \&
Titarchuk (CT95), the
post-shock region is comprised of a mixture of the Keplerian and
sub-Keplerian components.
Thus, though the sub-Keplerian halo (pre-shock) is optically thin for
the radiations from 
Keplerian disc, the post-shock torus could be optically thin, intermediate,
or even
thick depending on the Keplerian and sub-Keplerian rates
(see CT95, CTKE 96, C97 for details).
This is because (1) the Keplerian radiation falls on it at a glancing angle,
thereby increasing the
path length and (2) the mixed matter in this region has higher density.
CT95 showed soft radiation from the cool Keplerian Disc
are inverse-Comptonized by the CENBOL to produce the hard
radiation. If the sub-Keplerian halo rate (${\dot M}_h$)
is higher, then it supplies more hot electrons to the CENBOL than the soft
photons from the Keplerian disc and hence the soft photons cannot cool down
the
CENBOL significantly.  
Thus CENBOL remains puffed up and hot and can intercept a large number
of soft photons and inverse-Comptonize them to produce the hard power-law 
tail of the accretion disc spectrum --- a state called hard state.
If, on the other hand, the Keplerian accretion rate (${\dot M}_K$)
is higher, then it supplies more soft photons to cool down the 
CENBOL region. This results in more power to the soft end of the
accretion disc
spectrum --- a state known as the soft state. 
Recently, however,  Chakrabarti \& Mandal (2003) showed that raising 
the Keplerian rate even higher not necessarily softens the spectrum, for, the
Keplerian flow also adds to the number density of electrons in the post-shock
region and at some point, the spectrum starts to be hardened once more.

\begin {figure}
\vbox{
\vskip -0.0cm
\hskip 0.0cm
\centerline{
\psfig{figure=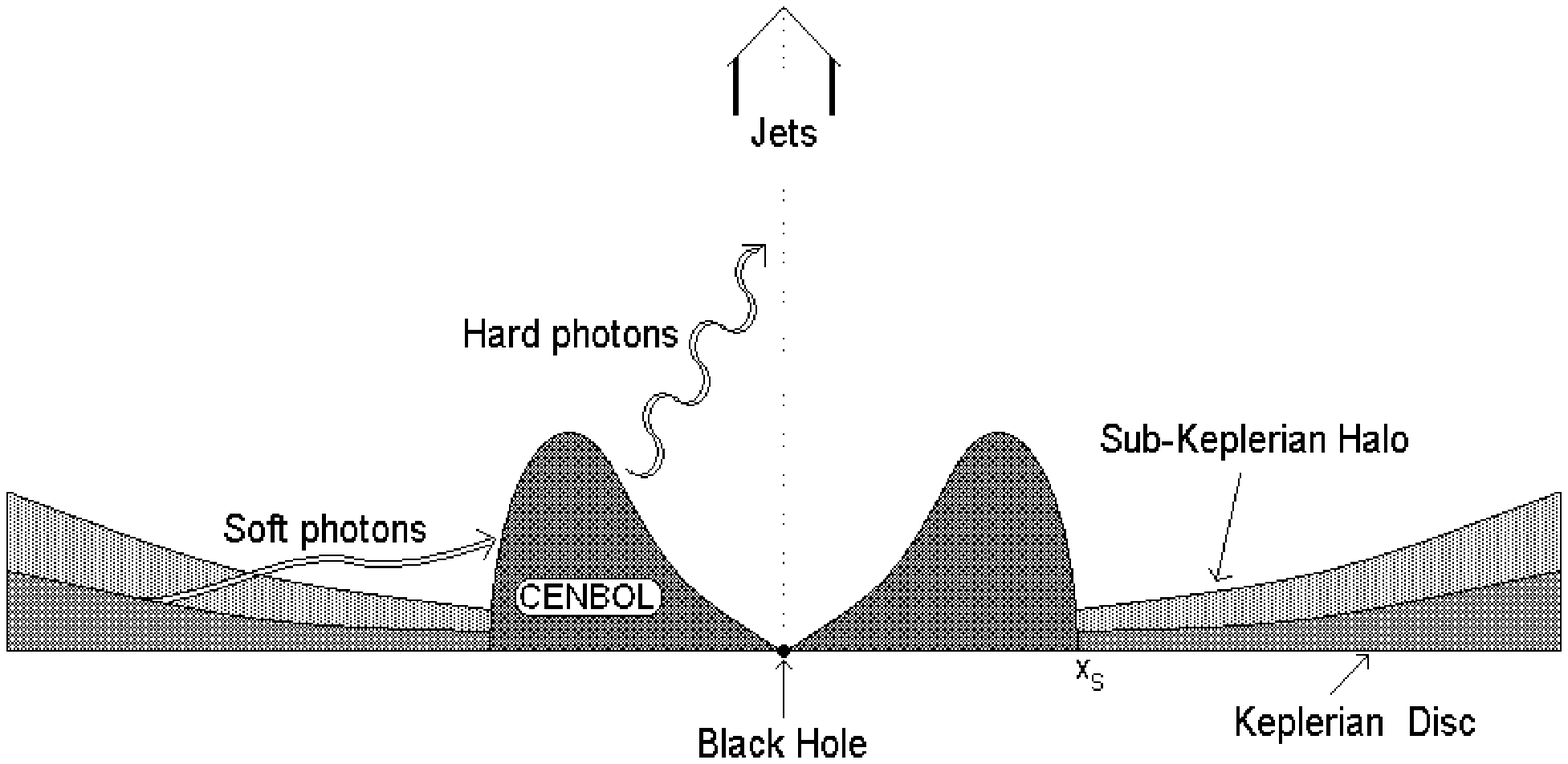,height=8truecm,width=15truecm}}}
\end{figure}
\begin{figure}
\vspace{-0.0cm}
\caption[] {Cross-sectional view of Two Component Accretion Disc Model. Only
the top half is shown.}
\end{figure}

This kind of hybrid
disc structure is known as the Two Component Accretion Flow or the TCAF disc
(see, CT95, CTKE 96, Ebisawa \etal 1996, C97),
and has wide observational support (Smith \etal 2001, 2002).
In Fig. (1), a schematic diagram of such a disc is presented. The Figure
shows the Keplerian disc,
is sandwiched by the sub-Keplerian halo. The shock location 
($x_s$) and the compact object are also shown. Jets are shown 
close to the axis of symmetry. Thus jets are illuminated by 
the Keplerian disc with soft photons and by the CENBOL
with hard photons.

Chakrabarti and his collaborators have also showed that the CENBOL
can drive a part of the infalling matter along the axis of symmetry to
form jets
(Chakrabarti, 1998; Chakrabarti, 1999; Das \& Chakrabarti, 1999;
Das \etal 2001).
There are wide support that the jets are indeed coming out from a region
within $50-100$ Schwarzschild radius of the black hole (Junor, Biretta \&
Livio, 1999).
Similarly, it is believed that jets are produced only in hard states 
(see, Gallo, Fender \& Pooley, 2003, and references therein).
Thus it is natural to study interaction of hard radiation from the
CENBOL and the outflowing jets, with the particular interest of studying,
whether momentum deposited to the jet material by these hard photons can
accelerate them to ultra-relativistic speeds. 

Investigation of interaction of radiation and astrophysical flow is not new. 
A number of astrophysicists have directed their
efforts in this particular field of study while the consideration of 
the associated accretion disc
depended on their personal choice or the popularity of the particular model
of accretion at the given time.
Icke (1980) studied the effect of radiative acceleration
of the gas flow above a Keplerian disc. But the
effect of radiation pressure on the gas flow was ignored.
Sikora and Wilson (1981) showed that even if the jet is collimated by
geometrically thick discs (Lynden-Bell, 1978; Abramowicz and Piran, 1980), 
radiation drag is important for astrophysical jets.
Piran (1982) while calculating the radiative acceleration of outflows
about the rotation axis of thick accretion discs, found out that in order to
accelerate outflows to ${\gamma}>1.5$ (where ${\gamma}$ is the
bulk Lorentz factor), the funnels must be short and steep, but such funnels
are found to be unstable.
Sol et. al. (1989) proposed a two-flow model for jets, one consists of
relativistic particles (electrons and positrons) and of relativistic
Lorentz factor, while the other being normal, mildly relativistic plasma.
In a very important paper,
Icke (1989) considered blobby jets about the axis of symmetry of
thin discs and he obtained the `magic speed'
of $v_{m}=0.451c$ ($c$ --- the velocity of light), $v_m$being the
upper limit of terminal speed.
Fukue (1996) extended this study for rotating flow above a thin
disc and drew similar conclusions, although for rotating flow, away
from axis of symmetry the terminal speed was found out to be a little
less than the magic speed of Icke.
To summarize, earlier works showed that it is not possible
to accelerate jets to ultra relativistic terminal velocities,
by radiations from the earlier accretion disc models.
What is more discouraging is the existence of moderate levels
of equilibrium speed ($v_{eq}$ \ie speeds above which there would be
radiative deceleration).
Recently Fukue \etal (2001) has done similar investigations on interaction
of radiation and pair plasma jets, but they took a disc model
which consisted of inner ADAF region (non luminous)
and outer slim disc (luminous), which resulted in a relativistic terminal
speed.

We are working in a different regime, \ie less luminous Keplerian disc
and more luminous post-shock torus or CENBOL.
Since hard radiations are expected to emerge out of
the optically thin CENBOL, its intensity (photon counts per unit area
per unit time) is low, 
but it `looks' directly into the jet vertically above and hence eventually
deposit its momentum into the later. Radiation from a hot CENBOL is 
likely a source of pair-production and hence the
possibility of radiative momentum deposition
is likely to be higher even  for radiations from CENBOL
hitting the outflow at an angle
(see, e.g., Yamasaki, Takahara, and Kusunose, 1999 
for the mechanism of pair-production from hot accretion flows.).
This is why we believe that the
direct deposition of momentum may be important.
Chattopadhyay \& Chakrabarti (2002a) showed
that hard radiations from the post-shock region (CENBOL) do accelerate
electron-proton plasma to mildly relativistic terminal speeds.
Chattopadhyay {\&} Chakrabarti (2002b) also reported that hard
radiations from
the CENBOL do not impose any upper limit for terminal speed. 

In the present paper, we solve the equations of photo-hydrodynamics
of jets for radiations
coming out from the TCAF discs, where the radiation fields from both
the inner
CENBOL and the outer Keplerian thin disc are considered. 
We show that, while the equilibrium velocity closer to the black hole
depends on $x_s$,
and the ratio between Keplerian disc and CENBOL luminosities, 
the terminal speed or the jet velocity at infinity depends on the
relative proportions and also on the actual magnitude of various moments of 
radiation.  We also show that, in hard states (in our parlance, CENBOL 
radiation dominating over Keplerian radiation), optically thin jets can 
be accelerated to ultra-relativistic speed.

In the next Section, we present the model assumptions and the equations
of motion and compute various moments of radiation field. 
In \S 3, we present our solutions and finally in \S 4, we draw our conclusions.  

\section{\bf Assumptions, Governing Equations and Computation
of the Moments of Radiation Field}

\subsection{Assumptions and Governing Equations}

In our analysis, the curvature effects due to the presence of the central
black hole mass is ignored. The metric is given by,
$ds^2=c^2dt^2-dr^2-r^2d{\phi}^2-dz^2$,
where, $r$, ${\phi}$, and $z$ are the usual coordinates in cylindrical geometry
and $ds$ is the metric in four-space. The four-velocities are $u^{\mu}$.
We follow the convention where the Greek indices signify all four components
and the Latin indices represent only the spatial ones.
The black hole is assumed to be non-rotating and hence the strong
gravity is taken care of by the so-called Paczy\'nski-Wiita potential
(\eg Paczy\'nski {\&} Wiita, 1980).
 
We also do not consider generation mechanism of jets.  As the
astrophysical jets are observed to be extremely collimated
(Bridle \& Perley, 1984), and generally aligned along the normal to the 
host galaxy, we assume the jet to be along the axis of symmetry. 
Thus the transverse
structure of the jet is ignored, \ie $u^r=u^{\phi}=0$ and 
${\partial}/{\partial}r={\partial}/{\partial}{\phi}=0$, where
$u^r$ and $u^{\phi}$ are radial and azimuthal components of
four velocity.
We are looking for steady state solutions. Hence, ${\partial}/{\partial}t=0$.
We also assume 
the gas pressure is negligible compared to the radiation pressure. This 
is perhaps the case especially inside the funnel wall close to the axis.
The derivation of the equations of motion of radiation hydrodynamics
for optically thin plasma, was investigated by a number of workers. A detailed
account of such derivation has been presented by Mihalas {\&} Mihalas (1984;
hereafter MM84) and Kato \etal (1998; hereafter K98), and are not presented
here.
Enforcing the above assumptions,
the equation of motion presented in MM84 and K98, reduces to;
$$
u^z\frac{du^z}{dz}=-\frac{GM_{B}}{(z-2)^2}+\frac{{\sigma}_{T}}{m}
\left[{\gamma}\frac{F^z}{c}-{\gamma}^2Eu^z-u^zP^{zz}+u^z
(2{\gamma}u^z\frac{F^z}{c}-u^zu^zP^{zz}) \right],
$$
where, $u^z$ is the $z$-component of four velocity, $G$, $M_{B}$,
${\sigma}_T$,
and $m$ are the universal gravitation constant, mass of the black hole,
Thomson scattering cross-section and
mass of the gas particle, respectively. $E$, $F^z$ and $P^{zz}$
are the radiative energy density, radiative flux and radiative pressure
on the axis of symmetry, and ${\gamma}({\gamma}=u_t)$ is the Lorentz factor.
The above equation can be re-written as,
$$
u^z\frac{du^z}{dz}=-\frac{GM_{B}}{(z-2)^2}+
\left[{\gamma}{\cal F}-{\gamma}^2{\cal E}u^z-u^z{\cal P}+u^z
(2{\gamma}u^z{\cal F}-u^zu^z{\cal P}) \right],
\eqno{(1)}
$$
where, ${\cal E}=\frac{{\sigma}_{T}}{m}E$, ${\cal F}=\frac{{\sigma}_{T}}
{mc}F^z$
and ${\cal P}=\frac{{\sigma}_{T}}{m}P^{zz}$.

For simplicity, we will not compute the shock location  $x_s$ or the
the CENBOL luminosity ($L_{C}$) -- instead, we will supply them as free
parameters. They can be easily computed from accretion parameters 
(\eg C89, CT95, Das \etal 2001, Chattopadhyay \etal 2003).
We assume that the outflow is made up of purely electron-positron pair
plasma. 

\subsection{Computation of radiative moments from TCAF disc}

The radiation reaching each point on the jet axis, is coming from two
parts of the disc, namely, the CENBOL and Keplerian disc, hence all
the radiative
moments should have both the contributions. 

\begin{figure}
\vbox{
\vskip 0.0cm
\hskip 0.0cm
\centerline{
\psfig{figure=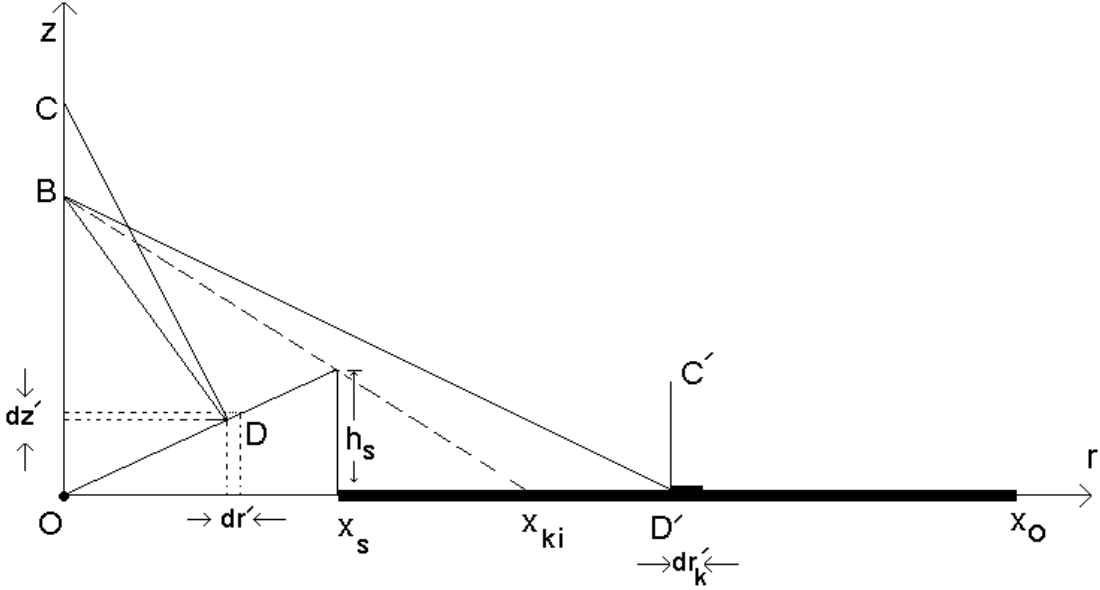,height=8truecm,width=15truecm}}}
\caption[]{Schematic diagram of a two component accretion flow (TCAF). 
$O$ is the position of the black hole. Centrifugal pressure dominated
boundary layer (CENBOL) is the puffed up region between $O$ and $x_s$,
 the shock location.  Thick line
between $x_s$ and $x_o$ is the Keplerian disc where $x_o$ is the outer
boundary of the Keplerian disc. ${\bf DC}$ and ${\bf D^{\prime}C^{\prime}}$
are the
local normals of the CENBOL and Keplerian disc. $B(0,z)$ is the field point
where various moments of radiation fields are computed. 
$D(r^{\prime},z^{\prime})$
is source point on the CENBOL and $D^{\prime}(r^{\prime}_{K},0)$ is the
source point on the Keplerian disc. $r^{\prime}$ is radial coordinate of the
CENBOL and $r^{\prime}_{K}$ is that of the Keplerian disc.
The sub-Keplerian halo is not shown.}
\end{figure}

In Fig. (2), a schematic
diagram of the cross-section of the disc structure is presented.
The black hole is situated at $O$.
The region bounded between $O$ and $x_s$ is the CENBOL, and the
thick line between $x_s$ and $x_o$ is the Keplerian disc, where $x_o$
is the outer boundary of the Keplerian disc. Thus $x_s$
is the outer boundary of CENBOL and inner boundary of the Keplerian
disc. The inner boundary of Keplerian disc, as seen from the position
$B$ is $x_{{K}i}$.
The shock height is $h_s{\sim}a_sx^{1/2}_s(x_s-1)$, where
$a_s$ is the equatorial sound speed at shock and depends on $x_s$.
In other words $h_s=h_s(x_s)$, but as we are not solving the accretion
disc equations simultaneously, we have to make some estimate of
$h_s$ which will closely mimic reality.
Chakrabarti solutions (C89, C90, CT95, C96) show that 
if $x_s=10r_g$ (where $r_g$ is the Schwarzschild radius),
then the temperature at the shock is
$T_{10}{\sim}1.56{\times}
10^{11}K$. Assuming that the shock temperature to be $T_s=T_{10}(10/x_s)$,
one can estimate the shock height to be $h_s{\sim}0.6(x_s-1)$. 
The inner surface of the CENBOL is assumed to be conical. 
The radiations from a point
$D(r^{\prime},z^{\prime})$ on the surface of CENBOL is primarily along
the local normal ${\bf DC}$. Similarly, the radiation from a point
$D^{\prime}(r^{\prime}_{K},0)$ on the Keplerian
disc, is along its local normal ${\bf D^{\prime}C^{\prime}}$.

The radiative moments at $B$ are;
$$
E=\frac{1}{c}\int Id{\Omega}=\frac{1}{c}\left({\int}_{C}
I_{C}d{\Omega}_{C}+{\int}_{K}I_{K}d{\Omega}_{K}
\right),
\eqno{(2a)}
$$
$$
\frac{F^i}{c}=\frac{1}{c}\int Il^id{\Omega}=\frac{1}{c}\left({\int}_{C}
I_{C}l^i_{C}d{\Omega}_{C}+{\int}_{K}I_{K}l^i_{K}d{\Omega}_{K} \right),
\eqno{(2b)}
$$
and
$$
P^{ij}=\frac{1}{c}\int Il^il^jd{\Omega}=\frac{1}{c}\left({\int}_{C}
I_{C}l^i_{C}l^j_{C}d{\Omega}_{C}+{\int}_{K}I_{K}l^i_{K}l^j_{K}d{\Omega}_{K}
\right).
\eqno{(2c)}
$$

In the Eqs. (2a-2c), $I$, $d{\Omega}$, $l^i$ are the frequency integrated 
intensity
from the disc, differential solid angle at $B$ and $l^i$ are the direction
cosines
at $B$, for example $l^z_{C}=(z-z^{\prime})/BD$ and
$l^z_{K}=z/BD^{\prime}$. Suffix $C$ and $K$ represent quantities 
linked to CENBOL and the
Keplerian disc respectively. The expressions of solid angles subtended at
$B$ from $D$ and $D^{\prime}$ are given respectively by, 
$$
d{\Omega}_{C}=\frac{r^{\prime}cosec{\theta}dr^{\prime}d{\phi}}
{{r^{\prime}}^2+(z-z^{\prime})^2}cos({\angle CDB}),
\eqno{(3a)}
$$
and
$$
d{\Omega}_{K}=\frac{r^{\prime}_{K}dr^{\prime}_{K}d{\phi}}
{{r^{\prime}_{K}}^2+z^2}cos({\angle C^{\prime}D^{\prime}B}),
\eqno{(3b)}
$$
where, ${\theta}$ is the semi-vertical angle of the inner surface of the CENBOL
which, for simplicity, is assumed to be constant.
It is to be noted that, from Fig. (2) and Eqs. (3a-3b), in general,
$l^z_{C}>l^z_{K}$ and for any unit differential area on the disc,
$d{\Omega}_{C}>d{\Omega}_{K}$. This implies that the contribution
from the CENBOL to the total radiation field moment is much more, compared to
that from the Keplerian disc. This will be clear when the comparative study 
of the moments are presented later [see Fig. (3)]. 

It is clear from Eq. (2b) that, due to the symmetry about the $z-$axis,
only non-zero pressure tensor components are $P^{ii}$ and
$F^r=F^{\phi}=0$ on the axis. Only $F^z{\neq}0$. 
As we consider jets on or about the 
axis of symmetry and the transverse structure is ignored, $P^{rr}$
and $P^{{\phi}{\phi}}$ do not enter the equation of motion as these
components are only coupled with $u^r(=0)$ and $u^{\phi}(=0)$, \ie
we have to only compute $P^{zz}$, which is exactly what is seen in Eq. (1).

From Fig. (2), it is clear that for the outflowing matter within the funnel,
\ie when $z<[h_sx_o/(x_o-x_s)]$, radiations from the Keplerian disc 
do not reach the electrons because these radiations are intercepted by CENBOL.
CENBOL will reprocess these intercepted photons,
and re-emit them in all directions, especially 
towards the jets because of especial geometry and directiveness
of the CENBOL surface. Due to shadowing effect mentioned above,
the radiation from the Keplerian disc reaching B is from $x_{{K}i}$ to $x_o$,
where $x_{{K}i}=x_sz/(z-h_s)$, with the additional 
constraints, $x_s{\leq}x_{{K}i}{\leq}x_o$ and $x_{{K}i}>0$.

As far as the CENBOL properties are concerned,
we follow CT95, where an effective temperature was
computed for the CENBOL and the radiation intensity 
was chosen to be uniform. That is:  
$I_{C}=L_{C}/{\pi}{\cal A}={\ell}L_{Edd}/{\pi}{\cal A}$= constant,
where $L_{C}$ and ${\cal A}$
are the CENBOL luminosity and the surface area of the CENBOL respectively.
$L_{Edd}$ is the Eddington luminosity and ${\ell}$ is the CENBOL luminosity
in units of $L_{Edd}$. The Keplerian disc intensity per unit solid angle is 
$I_{K}=\frac{3GM_{B}{\dot M}_{K}}{8{\pi}^2r^3_{K}}\left(1-
{\sqrt{\frac{3r_g}{r_{K}}}} \right)$ (NT73). 
Let us now multiply ${\sigma}_{T}/m$ with Eqs. (2a-2c), and then integrate
over the whole disc to obtain the following integrated quantities for the
moments,
\begin{eqnarray*}
\hspace{0.5cm} {\cal E} & = & {\cal E}_{{C}0}
\int^{x_s}_{r_{in}}\frac{2{\pi}
cot{\theta}{\{}r^{\prime}+(z-r^{\prime}cot{\theta})tan{\theta}{\}}r^{\prime}
dr{\prime}}{[(z-r^{\prime}cot{\theta})^2+r^{{\prime}2}]^{3/2}}
+{\cal E}_{{K}0} \int^{x_o}_{x_{{K}i}}\frac{2{\pi}z[
r^{{\prime}{-2}}_{K}-{\sqrt 3}r^{{\prime}{-5/2}}_{K}]
dr^{\prime}_{K}}
{(z^2+r^{{\prime}2}_{K})^{3/2}} \\
& = & {\cal E}_{{C}0}{\tilde E}_{\small C}(z,r_{in},x_s)+
{\cal E}_{{K}0}{\tilde{E}}_{\small K}(z,x_s,x_o) \\
& = & {\cal E}_{C}+{\cal E}_{K}, \hspace{12.3cm} (4a)
\end{eqnarray*}
where, $r_{in}$ is the inner boundary of the disc.
\begin{eqnarray*}
\hspace{0.5cm} {\cal F} & = & {\cal F}_{{C}0}
\int^{x_s}_{r_{in}}\frac{2{\pi}
cot{\theta}{\{}r^{\prime}+(z-r^{\prime}cot{\theta})tan{\theta}{\}}r^{\prime}
(z-r^{\prime}cot{\theta})
dr{\prime}}{[(z-r^{\prime}cot{\theta})^2+r^{{\prime}2}]^2} \\
& & +{\cal F}_{{K}0} \int^{x_o}_{x_{{K}i}}\frac{2{\pi}z[
r^{{\prime}{-2}}_{K}-{\sqrt 3}r^{{\prime}{-5/2}}_{K}]z
dr^{\prime}_{K}}
{(z^2+r^{{\prime}2}_{K})^2} \\
& = & {\cal F}_{{C}0}{\tilde F}_{\small C}(z,r_{in},x_s)+
{\cal F}_{{K}0}{\tilde{F}}_{\small K}(z,x_s,x_o) \\
& = & {\cal F}_{C}+{\cal F}_{K} \hspace{12.3cm} (4b)
\end{eqnarray*}

\begin{eqnarray*}
\hspace{0.5cm} {\cal P} & = & {\cal P}_{{C}0}
\int^{x_s}_{r_{in}}\frac{2{\pi}
cot{\theta}{\{}r^{\prime}+(z-r^{\prime}cot{\theta})tan{\theta}{\}}r^{\prime}
(z-r^{\prime}cot{\theta})^2
dr{\prime}}{[(z-r^{\prime}cot{\theta})^2+r^{{\prime}2}]^{5/2}} \\
& & +{\cal P}_{{K}0} \int^{x_o}_{x_{{K}i}}\frac{2{\pi}z[
r^{{\prime}{-2}}_{K}-{\sqrt 3}r^{{\prime}{-5/2}}_{K}]z^2
dr^{\prime}_{K}}
{(z^2+r^{{\prime}2}_{K})^{5/2}} \\
& = & {\cal P}_{{C}0}{\tilde P}_{\small C}(z,r_{in},x_s)+
{\cal P}_{{K}0}{\tilde{P}}_{\small K}(z,x_s,x_o) \\
& = & {\cal P}_{C}+{\cal P}_{K} \hspace{12.3cm} (4c)
\end{eqnarray*}

Constancy of $I_{C}$ allows us to have analytical expressions
for ${\tilde E}_{C}$, ${\tilde F}_{C}$ and ${\tilde P}_{C}$.
These were computed by Chattopadhyay {\&} Chakrabarti (2000, 2002ab)
and are not repeated here. We choose the unit of length  to be
$2GM_{B}/c^2$ --- the Schwarzschild
radius ($r_g$), unit of time to be $2GM_{B}/c^3$ and $M_{B}$ is the
unit of mass. Thus the unit of velocity is $c$. In such units,
constants in Eqs. (4a-4c) are,
$$
{\cal E}_{{\small C}0}={\cal F}_{{\small C}0}
={\cal P}_{{\small C}0}
=\frac{1.3{\times}10^{38}{\ell}{\sigma}_{T}}{2{\pi}cm{\cal A}GM_{\odot}}
\eqno{(5a)}
$$
and
$$
{\cal E}_{{\small K}0}={\cal F}_{{\small K}0}
={\cal P}_{{\small K}0}
=\frac{4.32{\times}10^{17}{\dot m_k}{\sigma}_Tc}{32{\pi}^2mGM_{\odot}}.
\eqno{(5b)}
$$
We have written the Keplerian accretion rate in units of the Eddington
accretion
rate ${\dot M}_{Edd}$, \ie ${\dot m}_{K} = {\dot M}_{K}/{\dot M}_{Edd}$.
It is to be noted that, two of the 
disc parameters \ie $r_{in}=1.5r_g$ and $x_o=1000r_g$ are kept constant 
through out the paper. In Fig. (3), the space dependent part of the 
moments ${\tilde E}$, ${\tilde F}$ and ${\tilde P}$
are compared for two shock locations.
In Figs. 3(a-b), the contributions from the CENBOL and the Keplerian disk
are plotted for $x_s=10r_g$ and in Figs. 3(c-d) the contributions 
from the CENBOL and the Keplerian disc are plotted, where 
the shock location is $x_s=20r_g$.
We see that as $x_s$ is increased, the CENBOL contributions increases
while the Keplerian contributions decrease.
We also see another fact that generally, in the hard state,
the CENBOL contribution to the total radiative moments dominates
over Keplerian counterparts.
We see from Fig. (3a) and (3c), that within the funnel (\ie $z{\leq}
h_s$), ${\tilde E}_{C}>{\tilde F}_{C}>{\tilde P}_{C}$.
We also see that, because of the relatively small size of the CENBOL,
${\tilde E}_{C}{\approx}{\tilde F}_{C}{\approx}
{\tilde P}_{C}$
for smaller values of $z$ [\eg $z{\sim}50.5r_g$ for Fig. (3a) and $z{\sim}
92.5r_g$
for Fig. (3c)],
while we see ${\tilde E}_{K}{\approx}{\tilde F}_{K}{\approx}
{\tilde P}_{K}$ at much larger distances [\eg $z{\sim}900r_g$ for Fig (3b)
and
$z{\sim}1500r_g$ for Fig. (3d)]. The small size of CENBOL ensures that the
direction 
cosines $l^z_{C}{\rightarrow}1$ for smaller $z$. In contrast, the larger 
size of the Keplerian disc ensures $l^z_{K}{\sim}1$ only for large $z$.
From Figs. 3(a-d), it appears that as ${\tilde E}_C {\gg}{\tilde E}_K$,
${\tilde F}_{C}{\gg}{\tilde F}_K$, and ${\tilde P}_{C}{\gg}{\tilde P}_K$.
So, if CENBOL is more
luminous, then CENBOL contributions would dominate the Keplerian
ones even at large $z$.
While ${\cal E}_{K0}$ ($={\cal F}_{K0}={\cal P}_{K0}$)
depends only on ${\dot m}_K$, 
${\cal E}_{C0}$ ($={\cal F}_{C0}={\cal P}_{C0}$),
depends on both ${\ell}$ and the CENBOL surface area.
From Figs. (3a) and (3c), we see that with the increase of $x_s$, 
even if the radiative moments from CENBOL, say for example,
${\tilde E}_C$ increases moderately, but calculations show that
${\tilde E}_{C0}$ decreases appreciably with $x_s$ (\eg ${\tilde E}_{C0}$
for $x_s=10r_g$ is about four times larger than ${\tilde E}_{C0}$
for $x_s=20r_g$).
So Keplerian contributions might be comparable to that
due to the CENBOL at large $z$, depending on specific combinations
of all the parameters
${\ell}$, ${\dot m}_K$ and $x_s$, especially for large values of
$x_s$ and low ${\ell}$. 
We also observe that the shadow effect of CENBOL ensures that at 
$z{\leq}(h_sx_o)/(x_o-x_s)$,
${\tilde E}_{K}={\tilde F}_{K}={\tilde P}_{K}=0$.

\begin {figure}
\vbox{
\vskip -0.0cm
\hskip 0.0cm,  
\centerline{
\psfig{figure=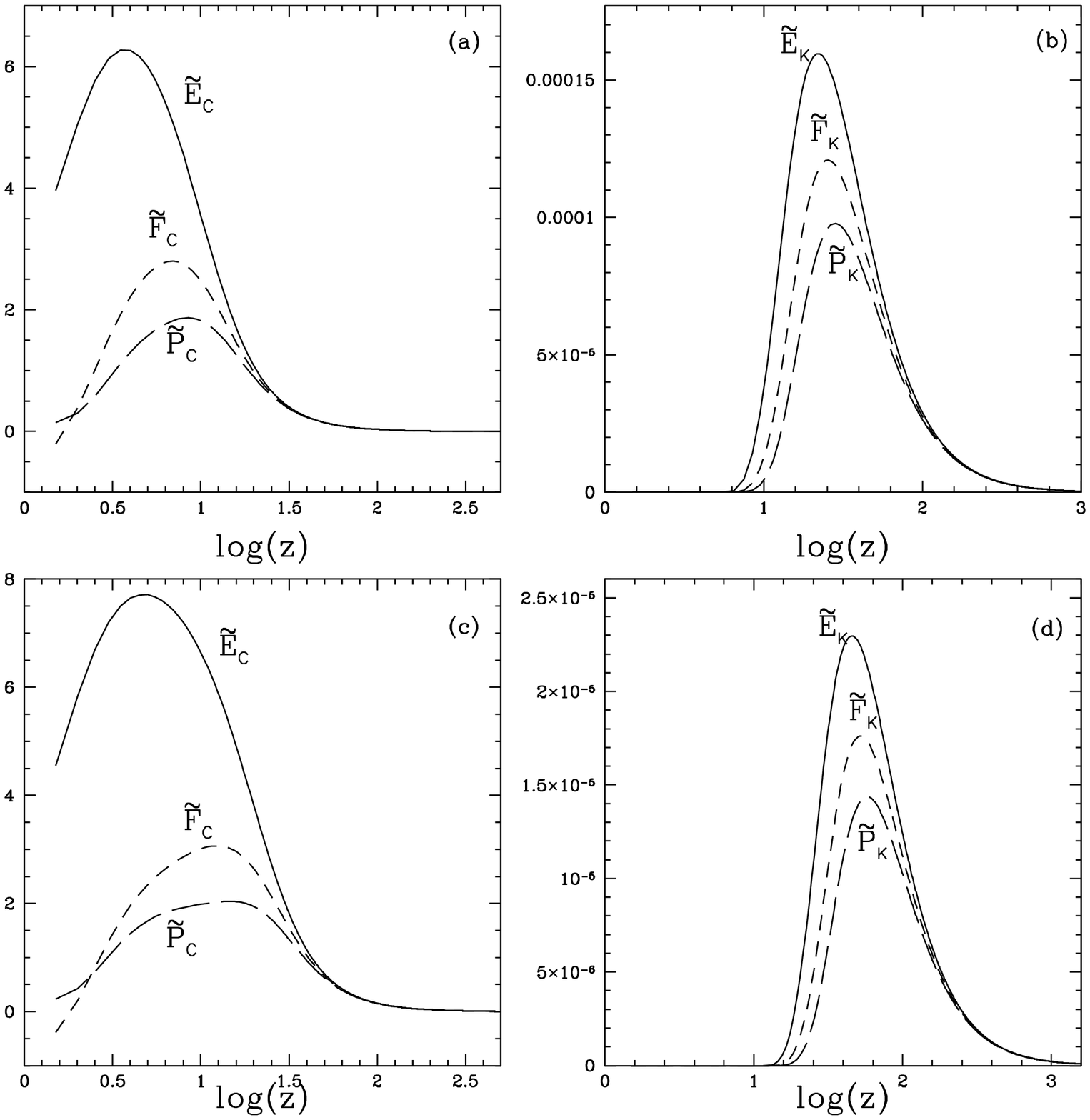,height=16truecm,width=15truecm}}}
\end{figure}
\begin{figure}
\vspace{0.0cm}
\caption[] {Variations of spatial part of radiation moments, ${\tilde E}$
(solid)
${\tilde F}$ (dashed) and ${\tilde P}$ (long-dashed) with log(z). (a)
Contributions
from the CENBOL denoted by suffix $C$, and (b) Contributions from the
Keplerian
disc denoted by suffix $K$, shock location is $x_s=10r_g$. (c)
Contributions
from the CENBOL for $x_s=20r_g$, and (d) Contributions from the
Keplerian disc
for $x_s=20r_g$. The suffixes have the same meaning as in the previous
two Figures.}
\end{figure}

\section{Radiative acceleration}

In the rest of the paper, we will use geometrical units defined in the last section,
but for simplicity we will keep the same symbols representing variables
as in Eq. (1).
We define a three-velocity $v$ such that $v^2=-u_iu^i/u_tu^t=-u_zu^z/u_tu^t$.
Thus $u^z={\gamma}v$ and $u_z=-{\gamma}v$. Under such considerations Eq. (1)
takes the form,
$$
\frac{dv}{dz}=\frac{-\frac{1}{2(z-1)^2}+[{\gamma}{\cal F}-{\gamma}^3v{\cal E}
-{\gamma}v{\cal P}+{\gamma}^3(2v^2{\cal F}-v^3{\cal P})]}
{{\gamma}^4v}. 
\eqno{(6)}
$$
The first term in the right-hand side of Eq. (6) is the gravitational 
term with dimensionless Pacz\'ynski-Wiita
potential and the term in the square bracket is the radiative
acceleration term. We notice that the radiative acceleration term depends both
on $v$ and the radiative moments. The first and fourth term in the square 
bracket are accelerating terms while the second, third and fifth 
terms are the decelerating terms, collectively known as the radiation drag terms.
We also see that there is a ${\gamma}^4$ term in the denominator of the
r. h. s of Eq. (6). This term ensures that, apart from radiation drag,
outflowing matter will be slowed down as $v{\rightarrow}1$.

\subsection{Equilibrium velocity}

Let us now discuss the concept of equilibrium velocity.
As a concept, equilibrium velocity is not a new one and has been
extensively discussed by a number of astrophysicists
(see, Fukue 2003, and references there in for details). We want to study
this issue in the context of jets in the radiation field of TCAF discs.
It is defined in a manner that at $v=v_{eq}$ the square bracket term 
in Eq. (6) is zero. Thus, for $v>v_{eq}$ there is 
deceleration and for $v<v_{eq}$ there is radiative
acceleration.

Putting the square bracket term in Eq. (6) to zero we have,
$$
{\cal F}v^2_{eq}-({\cal E}+{\cal P})v_{eq}+{\cal F}=0.
\eqno{(7)}
$$

This is a quadratic equation of $v_{eq}$ whose solution is,
\begin{eqnarray*}
\hspace{4.0cm} v_{eq}(z) & = & \frac{({\cal E}+{\cal P})-
{\sqrt{({\cal E}+{\cal P})^2-4{\cal F}^2}}}{2{\cal F}} \\
& = & {\xi}-{\sqrt{({\xi}^2-1)}}, \hspace{7.2cm} (8)
\end{eqnarray*}
where ${\xi}=({\cal E}+{\cal P})/2{\cal F}$.
From Eqs. (4a-5b), we have,
$$
{\xi}(z)=\frac{({\tilde E}_{C}+{\tilde P}_{C})
+\frac{{\dot m}_{K}}
{\ell}{\zeta}({\tilde E}_{K}+{\tilde P}_{K})}
{2({\tilde F}_{C}+
\frac{{\dot m}_{K}}{\ell}{\zeta}{\tilde F}_{K})},
\eqno{(9)}
$$
where, ${\zeta}=6.6{\times}10^{-13}c^2{\cal A}$. It is to be noted
that ${\xi}$ depends on $x_s$ as well as on the ratio 
${\dot m}_{K}/{\ell}$,
but not separately ${\dot m}_{K}$ and ${\ell}$.
In case $({\dot m}_K{\zeta})/{\ell} {\ll} 1$, 
$v_{eq}$ is completely determined by ${\tilde E}_{C}$, ${\tilde F}_{C}$
and ${\tilde P}_{C}$, and ${\ell}$ has no effect in determining $v_{eq}$.
We also see that
${\xi}$ does not depend on the mass of the gas particles $m$. Thus
${\xi}$ (and thus $v_{eq}$) is the same for both electron-proton 
plasma as well electron-positron plasma, provided $x_s$ and ${\dot m}_{K}/{\ell}$ is
the same in both the cases. It is clear from Eqs. (8-9), that as 
${\cal E}{\approx}{\cal F}{\approx}{\cal P}$, $v_{eq}{\rightarrow}1$. Therefore
if the CENBOL contribution dominates then, $v_{eq}{\sim}1$ within few tens
of Schwarzschild radii above the disc plane. If Keplerian
radiation dominates, then the condition is achieved at much larger distance,
as we shall see later.
We also see that no outflow is possible \ie $v_{eq}{\leq}0$, i.e., if

$$
{\cal F}{\leq}0.
\eqno{(10)}
$$

From Fig. (3) we see that very close to the horizon, due to the
torus geometry of the CENBOL
${\cal F}<0$, hence very close to the black hole,
not only enormous gravitational pull but also the
radiative force pushes matter inward. 

\noindent {\it Icke's magic speed}: \\
Let us now recover an important result from Eq. (8). If we consider
a thin-Keplerian disc  of infinite size, then $P^{zz}_{K}=P^{rr}_{K}=
P^{{\phi}{\phi}}_{K}=\frac{1}{3}E_{K}$ and $F^z_{K}=\frac{c}{2}E_{K}$,
and there is no CENBOL in this particular case. Under such conditions
${\cal E}=2{\cal F}=3{\cal P}$, putting these in Eq. (8) we have,
$$
v_{eq}=v_m=\frac{1}{3}(4-{\sqrt {7}})=0.451
{\equiv} \mbox{ magic speed of Icke }!
$$
Thus we see that, if the radiation field, \ie the radiative properties 
of the disc, can be prescribed, then $v_{eq}$ can be found out easily.

\begin {figure}
\vbox{
\vskip -0.0cm
\hskip 0.0cm
\centerline{
\psfig{figure=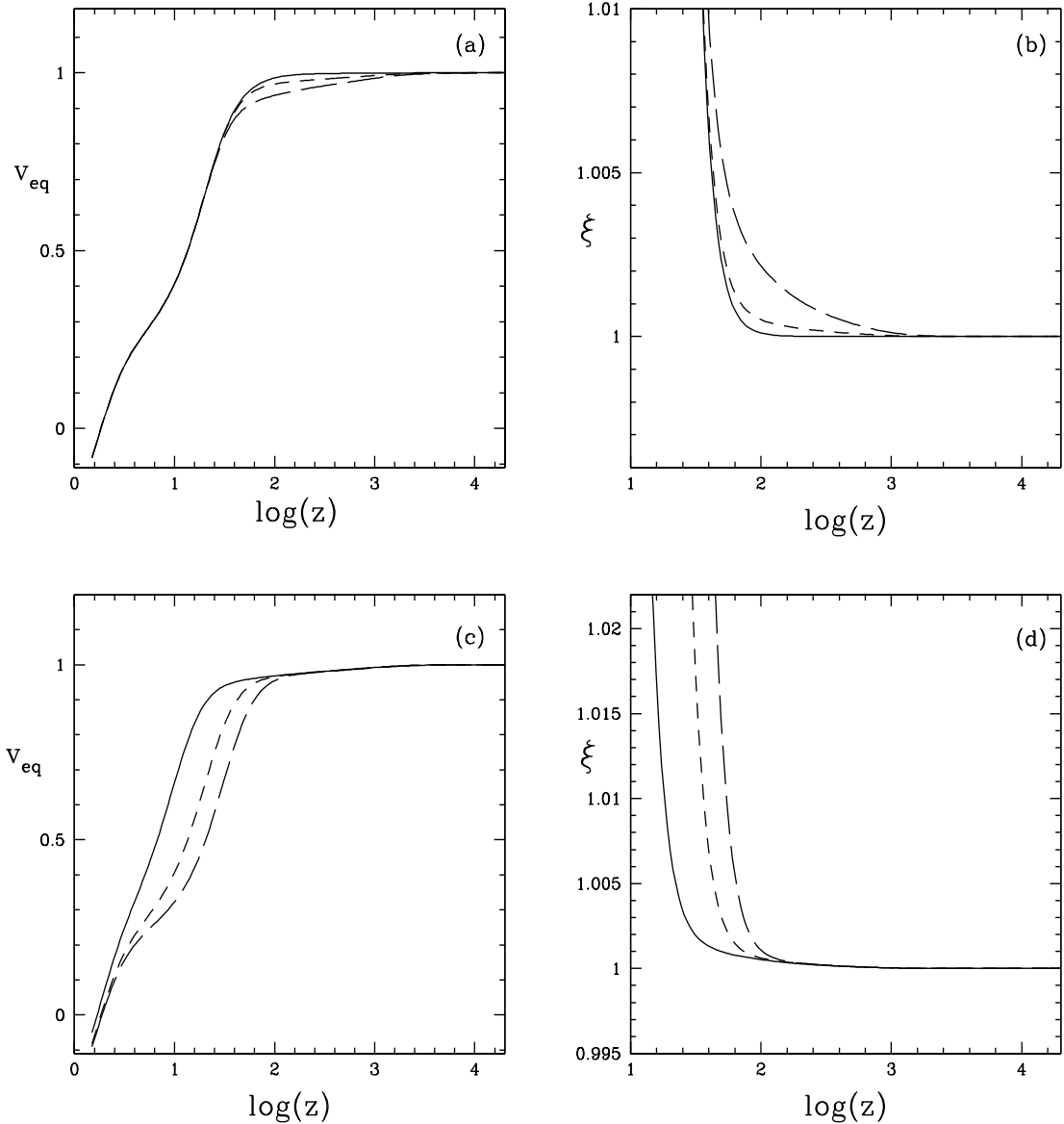,height=16truecm,width=15truecm}}}
\caption[] {Variation of (a) $v_{eq}$ and (b) ${\xi}$ with $log(z)$ for
${\ell}=0.3$ and $x_s=20r_g$.
Various curves correspond to ${\dot m}_{K}=0.01$ (solid), ${\dot m}_{K}=1$
(dashed) and ${\dot m}_{K}=6$
(long-dashed).
Variation of (c) $v_{eq}$ and (d) ${\xi}$ with $log(z)$.
for ${\ell}=0.12$ and ${\dot m}_K=0.5$.
Various curves corresponds $x_s=10r_g$ (solid), $x_s=20r_g$ (dashed)
and $x_s=30r_g$ (long-dashed). }
\end{figure}

\noindent {\it Equilibrium speed from a TCAF disc}:  \\
Let us now concentrate on the radiations from TCAF discs. As discussed in the
introduction, jets are produced in the hard state of the accretion disc  when the
CENBOL is hotter
and hard state means more power is on the high energy end of the spectrum. 
That is, $L_{C}>L_{K}$. The Keplerian disc luminosity is given by $L_{K}=r^2_g
{\int}^{x_o}_{x_s}2{\pi}I_{K}2{\pi}r_{K}dr_{K}$,
which is a function of ${\dot m}_{K}$, $x_s$ and $x_o$, which when
integrated can be expressed as,
$$
L_{K}=r^2_g{\int}^{x_o}_{x_s}2{\pi}I_{K}2{\pi}r_{K}dr_{K}
=\frac{3}{4}{\dot m}_{K}\left [-\frac{1}{r_{K}}+\frac{2}{3r_{K}}
{\sqrt{\frac{3}{r_{K}}}} \right]^{x_o}_{x_s}L_{Edd}.
\eqno{(11)}
$$
Thus the Keplerian disk luminosity in units of the Eddington luminosity
can be defined as ${\ell}_K=L_{K}/L_{Edd}$.
As has been stated before, presently we do not compute $L_{C}$, 
but supply it. Typical values of $\ell$ that we shall employ 
should, in general, depend on $\ell_K$ itself because
$\ell \sim  \Lambda \ell_K$, where $\Lambda$ is the enhancement factor by
which
incident photon intensity is increased due to Comptonization and has a
value of around $20-30$
in hard states (CT95). 
Thus, for instance, if $\ell_K \sim 0.05$, a typical value of 
$\ell \sim 0.1-0.15$. However, while we shall choose $\ell$ and $\ell_K$
in these
regions, we shall use them as free parameters, since presently we are not
interested in computing the
terminal speed as a function of the spectral slope $\alpha$, though,
strictly speaking, 
it would be a function of $\alpha$.

In our case, we see that, there are two sources of radiation from the disc,
(i) the CENBOL and (ii) the Keplerian disc.
Let us first concentrate on the CENBOL contribution.
From Figs. (3a) and (3c), we see that for $z{\rightarrow} {\mbox{ few }}
{\times}10r_g$,
$({\cal E}_{C}+{\cal P}_{C}){\sim}2{\cal F}_{C}$, hence Eq. (8)
determines $v_{eq}{\sim}1$ at such distances.

In general, in the hard state also, $L_{K}$ is not negligible. Though
at $z{\rightarrow}$large, ${\tilde E}_K{\approx}{\tilde F}_K
{\approx}{\tilde P}_K$, but at $z{\sim}{\mbox{ few }}{\times}10r_g$,
$({\cal E}_{K}+{\cal P}_{K})>2{\cal F}_{K}$. Hence for higher
Keplerian luminosity, $v_{eq}{\sim}1$ is achieved at distances
around a thousand Schwarzschild radii.
In Figs. (4a) and (4b), we have plotted $v_{eq}$  and $\xi$ with $log(z)$.
Various curves correspond to ${\dot m}_{K} \sim 0.01$ (solid),
${\dot m}_{K}=1$(dashed) and ${\dot m}_{K}=6$(long-dashed).
We choose ${\ell}=0.3$ and $x_s=20r_g$ for all the plots.
We see that close to the black hole, $v_{eq}$ is independent of ${\ell}_{K}$, 
because at such distances radiation from the CENBOL dominates. If 
$L_C{\gg}L_{K}$ (\ie solid curve),
then we see that $v_{eq}{\rightarrow}1$ at around $100r_g$. 
For ${\dot m}_{K}=1$
(${\equiv}{\ell}_{K}{\sim}0.03$, \ie dashed curve), 
and for ${\dot m}_{K}=6$ (\ie long-dashed curve),
$v_{eq}{\rightarrow}1$ at distances over $1000 r_g$. 
One should note that at $z<1.85r_g$, $v_{eq}<0$. The reason for this 
can be clearly seen from Fig. (3c), which shows that ${\cal F}<0$
at the same distance from the black hole.
This means that very close to the black hole not only
gravity pushes in matter but also radiation would also
push matter inside, thus vindicating Eq. (10). To clarify
this point we have plotted $\xi$
with $log(z)$ in Fig. (4b). Curve styles match those of the
corresponding case in Fig. (4a).
We see that within the funnel like region of the CENBOL, 
the radiation is completely
dominated by the CENBOL itself. And also due to this particular geometry
${\xi}>1$, resulting $v_{eq}$ much lesser than the velocity of light.
In case ${\dot m}_{K}{\rightarrow}$small (solid), for $z>100r_g$,
${\xi}{\rightarrow}1$, resulting $v_{eq}
{\rightarrow}1$. In case ${\dot m}_{K}{\rightarrow}$ large (dashed and
long-dashed curves),
only at around a thousand 
Schwarzschild radii, ${\xi}{\sim}1$. This means that, if one
increases
${\dot m}_K$, then the higher velocities achieved due to the
acceleration of jets
by CENBOL photons, might be decelerated. 
It is quite obvious though, that the effect of Keplerian radiation is
quite marginal.
Of course one should keep in mind that
if the radiative moments due to CENBOL and Keplerian radiation are
comparable
at infinite distances,
then there is a possibility of increased terminal speed, with
the increase in
${\dot m}_K$.

\begin {figure}
\vbox{
\vskip -0.0cm
\hskip 0.0cm
\centerline{
\psfig{figure=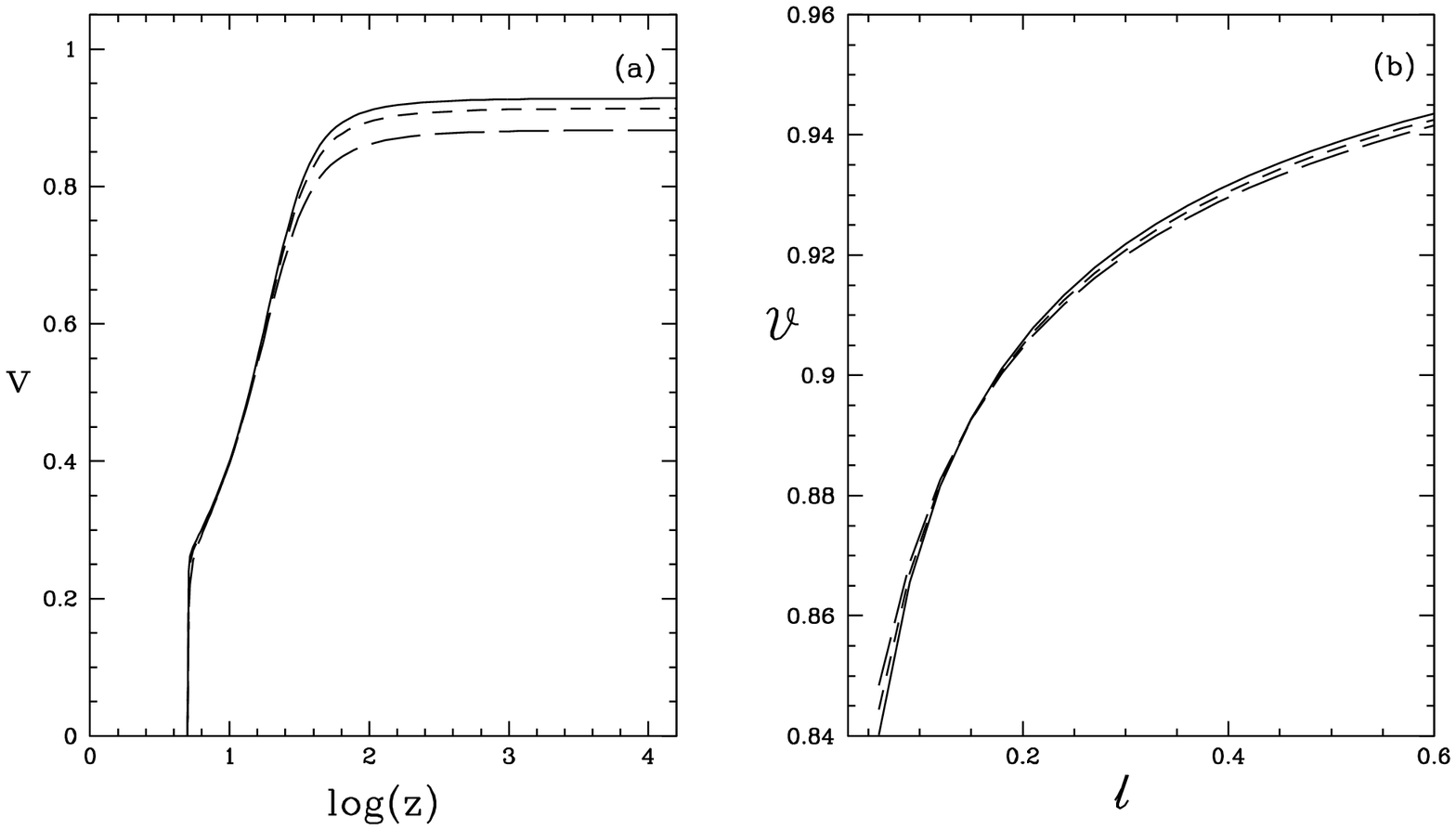,height=16truecm,width=15truecm}}}
\end{figure}
\begin{figure}
\vspace{-6.0cm}
\caption[] {(a) Variation of $v$ with $log(z)$, for ${\dot m}_K=0.5$.
Various curves correspond to ${\ell}=0.36$ (solid),
${\ell}=0.24$ (dashed) and ${\ell}=0.12$ (long-dashed).
(b) Variation of ${\vartheta}$ with ${\ell}$.
Various curves correspond to ${\dot m}_K=0.4$ (solid),
${\dot m}_K=1.4$ (dashed) and ${\dot m}_K=2.4$ (long-dashed). 
The shock location $x_s=20r_g$, for both the cases.
Injection parameters $z_{in}=5r_g$, $v_{in}=10^{-3}$.}
\end{figure}

We have seen from Eq. (9), that ${\xi}$ not only depends on ${\dot m}_K/{\ell}$
but also on $x_s$. We now investigate the dependence of $v_{eq}$ on $x_s$. 
In Fig. (4c), $v_{eq}$ is plotted with $log (z)$ for
${\ell}=0.12$ and ${\dot m_K}=0.5$. Various curves represent
$x_s=10r_g$ (solid), $x_s=20r_g$ (dashed) and $x_s=30r_g$ (long-dashed).
With the increase in $x_s$, as the size of CENBOL increases, hence
it can only behave like a point source (for which ${\xi}{\rightarrow}1$)
farther out, and we see that $v_{eq}{\rightarrow}1$ farther outwards
from the black hole. To clarify, we have also plotted
${\xi}$ with $log(z)$ for
${\ell}=0.12$ and ${\dot m_K}=0.5$ in Fig. (4d). Various curves represent
$x_s=10r_g$ (solid), $x_s=20r_g$ (dashed) and $x_s=30r_g$ (long-dashed).
As has been just explained, we see that ${\xi}{\rightarrow}1$ for larger
value of $z$, as $x_s$ is increased.

\subsection{Velocity profile}

So far, we have only discussed about the equilibrium velocity
and its dependence on ${\ell}$, ${\dot m}_{K}$ and $x_s$.
This was done to study the upper limit of allowed velocity
as a function of $z$.
We have seen that, if CENBOL radiation
dominates over that from the Keplerian disc, then jets
can be accelerated to very high velocities within around $100r_g$,
but $v_{eq}$ is
only a measure of velocity which signifies only the domain of
radiative deceleration or acceleration, and is not the actual velocity. 
From Eq. (6) it is quite clear that the radiative acceleration 
and hence $v$ itself, depends
on all three quantities ${\cal E}$, ${\cal F}$, ${\cal P}$ and not only
just on ${\xi}$. Equation (6) also shows, radiative acceleration
also depends on $v$ in a very complicated way.
To compute actual velocity, one has to integrate Eq. (6).
Jets are believed to be produced from the post shock region (Chakrabarti 1999,
Das \& Chakrabarti 1999, Das \etal 2001). If they are generated
from the post-shock region then they should start with very small velocity.
In this paper we are not concentrating on the generation of jets.
Thus we just put the injection height to be close to the black hole and
the injection velocity to be small.
We choose
injection parameters to be $z_{in}=5r_g$ and $v_{in}=10^{-3}$.
Before discussing the results, let us define terminal speed.
The terminal speed (${\vartheta}$) is the constant velocity at
infinite distances
or ${\vartheta}=v|_{z{\rightarrow}{\infty}}$.
In Fig. (5a), three velocity $v$ is plotted with $log(z)$, where
${\dot m}_K=0.5$ and $x_s=20r_g$. Various curves correspond to
${\ell}=0.36$
(solid), ${\ell}=0.24$ (dashed) and ${\ell}=0.12$
(long-dashed).
We see that close to the black hole as $v{\ll}1$, ${\cal F}$ dominates 
resulting in a steep rise in $v$, as $v$ increases the jet starts to feel
the drag force, resulting in a somewhat less steep increase in $v$.
At $z>100r_g$, the radiative moments tend to become weak and the jet
settles to a constant velocity at large $z$ or the terminal speed
${\vartheta}$.
It is clear that, $v$ increases with ${\ell}$ and in the three cases
depicted,
the terminal velocity of jets are ${\vartheta}{\sim}0.93$ (solid),
${\vartheta}{\sim}0.91$ (dashed) and ${\vartheta}{\sim}0.88$ (long-dashed).
It may seem curious that even if ${\ell}$ is increased by equal interval
the terminal speed achieved, does not increase by equal interval.
The reason is two fold, first of all from Eq. (6), one can see that
the gradient
of $v$ is a non-linear function of the moments. The second reason
being the
existence of ${\gamma}^4$ term in Eq. (6), which suppresses
the acceleration,
as $v$ increases to values close to that of light.

In Fig. (5b), the terminal speed ${\vartheta}$ is plotted with ${\ell}$, for
$x_s=20r_g$. Various curves correspond to ${\dot m}_K=0.4$ (solid),
${\dot m}_K=1.4$ (dashed) and ${\dot m}_K=2.4$ (long-dashed).
The terminal speed ${\vartheta}$ increases with ${\ell}$, though
for higher ${\ell}$, the increase of ${\vartheta}$ decreases, for the
same two reasons discussed for the previous Figure.
We also see that the dependence of ${\vartheta}$ on ${\dot m}_K$
is marginal. In Fig. (3), we have seen that the CENBOL contribution
to the various space dependent part of the radiative moments, is a
few orders of magnitude
higher than that from the Keplerian contribution, hence
the marginal dependence of ${\vartheta}$ with ${\dot m}_K$ is
expected. It is to be noted though, that for ${\ell}>0.15$, ${\vartheta}$
decreases with increasing ${\dot m}_K$
and for ${\ell}<0.15$, ${\vartheta}$ increases with increasing
${\dot m}_K$, albeit the dependence is very weak.
For the higher values of ${\ell}$, the jet, powered by the
CENBOL radiation, achieves very high velocity within a few
tens of Schwarzschild
radii. But for the next thousand of Schwarzschild radii or so,
${\xi}>1$, as $({\cal E}_K+{\cal P}_K)>2{\cal F}_K$ in that region
[see, Fig. (3c-3d)]. When $v$ becomes 
high, the drag force increases, and slows down the jet.
At a larger distance, a slight increase in radiative moments, due to the
increase in ${\dot m}_K$, is not sufficient to increase ${\vartheta}$. 
When ${\ell}$ is small, the velocity achieved within few tens of
Schwarzschild radii is not that high, hence the drag force is less.
But at large distances, the radiative moments due to CENBOL
and Keplerian disc becomes comparable hence there is a slight increase in 
${\vartheta}$ with ${\dot m}_K$.

\begin {figure}
\vbox{
\vskip -0.0cm
\hskip 0.0cm
\centerline{
\psfig{figure=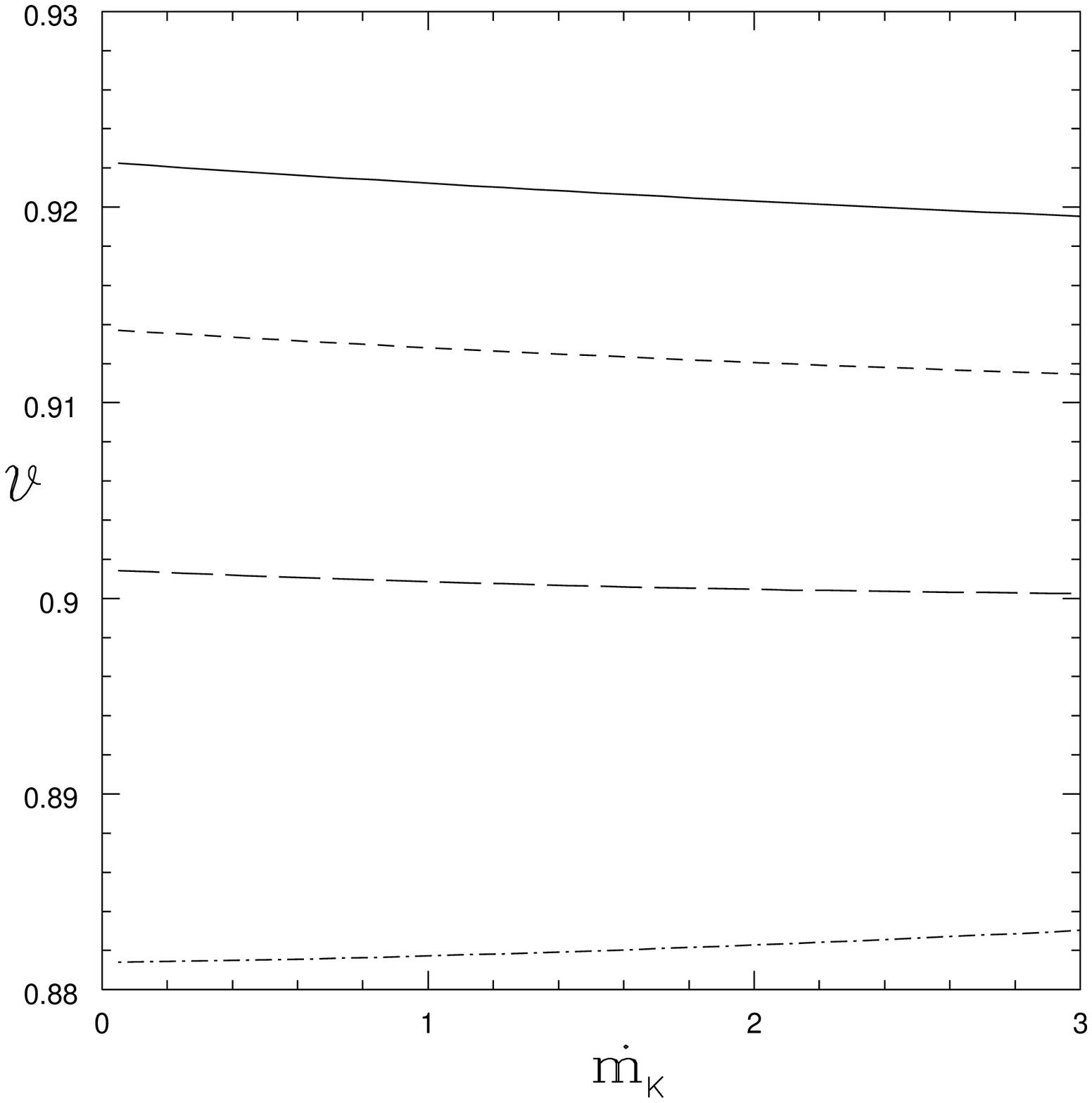,height=10truecm,width=10truecm}}}
\end{figure}
\begin{figure}
\vspace{-0.7cm}
\caption[] {Variation of ${\vartheta}$ with ${\dot m}_K$, for
$x_s=20r_g$. Various curves correspond to ${\ell}=0.3$ (solid),
${\ell}=0.24$ (dashed) and ${\ell}=0.18$
(long-dashed) and ${\ell}=0.12$ (dashed-dotted).
Injection parameters are $z_{in}=5r_g$, $v_{in}=10^{-3}$.}
\end{figure}

In Fig. (6), ${\vartheta}$ is plotted with ${\dot m}_K$, for $x_s=20r_g$.
Various curves correspond to ${\ell}=0.3$ (solid), ${\ell}=0.24$
(dashed), ${\ell}=0.18$ (long dashed),
and ${\ell}=0.12$ (dashed-dotted). We see that, as in Fig. (5b),
for higher ${\ell}$, ${\vartheta}$ decreases with increasing ${\dot m}_K$ 
(see,  solid, dashed and long dashed curves), but for lower
values of ${\ell}$, ${\vartheta}$ increases
with increasing ${\dot m}_K$ (see, dashed-dotted curve).
So one can conclude that in the hard state, jets are basically
accelerated by radiations from CENBOL and the terminal velocity
has a weak dependence on
radiation from Keplerian disc.

\begin {figure}
\vbox{
\vskip -0.0cm
\hskip 0.0cm
\centerline{
\psfig{figure=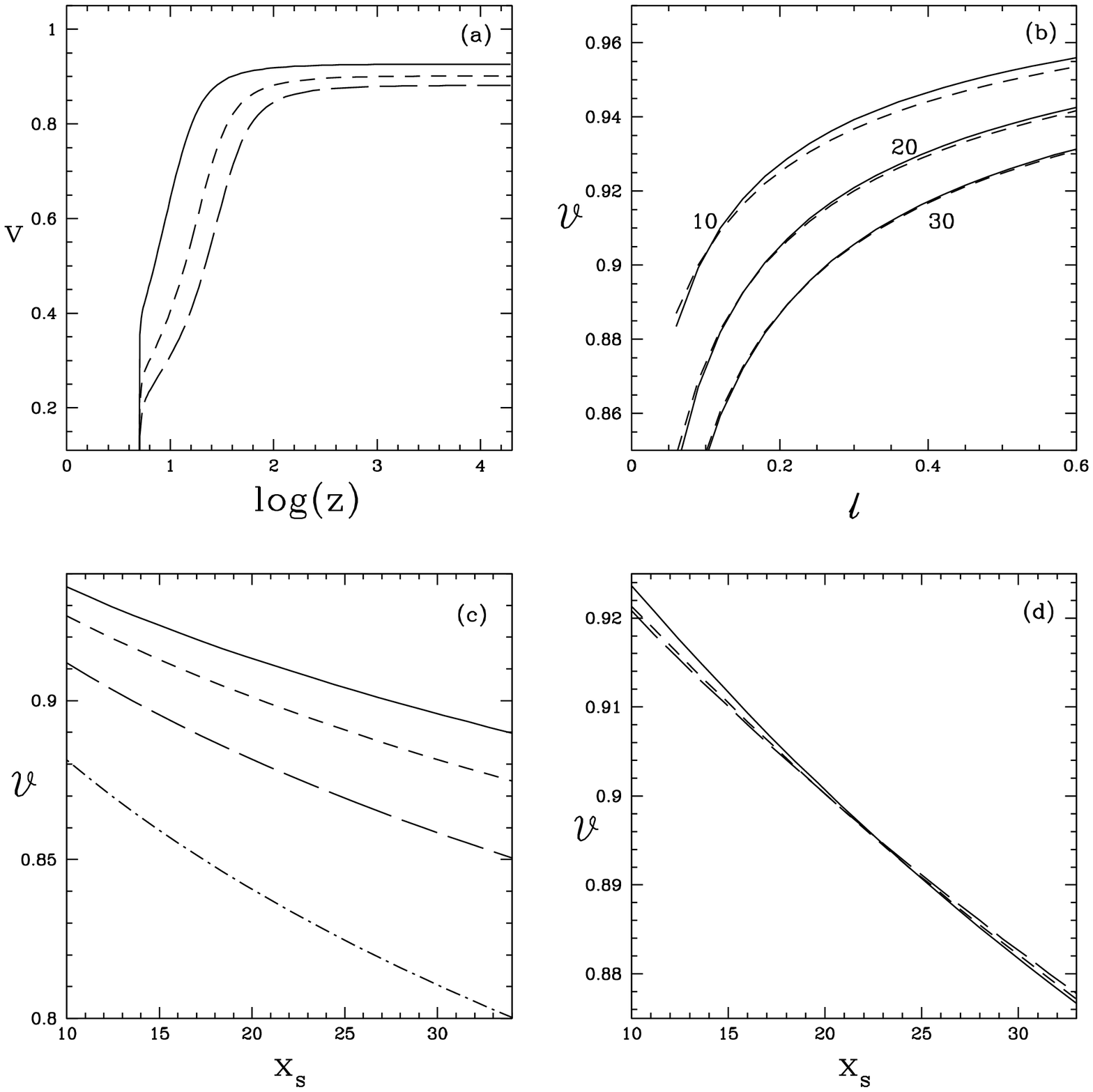,height=16truecm,width=15truecm}}}
\caption[] {(a) Variation of $v$ with $log(z)$ for
${\ell}=0.18$ and ${\dot m}_K=0.5$.
Various curves correspond to $x_s=10r_g$ (solid),
$x_s=20r_g$ (dashed) and $x_s=30r_g$
(long-dashed).
(b) Variation of ${\vartheta}$ with ${\ell}$.
The solid curves correspond to ${\dot m}_K=1.4$
and dashed curves correspond to ${\dot m}_K=2.4$.
The shock locations $x_s$, are marked on each pair of
solid and dashed curves ($x_s=10r_g, 20r_g, 30r_g)$.
(c) Variation of ${\vartheta}$ with $x_s$, for ${\dot m}_K=0.5$.
Various curves represent ${\ell}=0.24$ (solid), ${\ell}=0.18$
(dashed), ${\ell}=0.12$ (long dashed) and ${\ell}=0.06$ (dash-dot).
(d) Variation of ${\vartheta}$ with $x_s$, for ${\ell}=0.18$.
Various curves represent ${\dot m}_K=1.5$ (solid)
${\dot m}_K=3$ (dashed) and ${\dot m}_k=4.5$ (long dashed).
Injection parameters $z_{in}=5r_g$, $v_{in}=10^{-3}$.}
\end{figure}

Let us now investigate the dependence on $x_s$.
In Fig. (7a), $v$ is plotted with $log(z)$, for ${\ell}=0.18$ and
${\dot m}_K=0.5$. Various curves correspond to $x_s=10r_g$ (solid),
$x_s=20r_g$ (dashed) and $x_s=30r_g$ (long-dashed). We see that 
with the increase in $x_s$, the acceleration is reduced.
For $x_s=10r_g$ (solid),
we see that the jet is experiencing tremendous acceleration and
within a short distance of $z{\sim}40r_g$, it
achieves a velocity ${\sim}0.9$,
and gradually settles to a terminal value of ${\vartheta}=0.93$.
As the shock location is increased to $x_s=20r_g$ (dashed)
and $x_s=30r_g$ (long-dashed), acceleration is weaker and the terminal
speeds achieved are ${\vartheta}=0.9$ and ${\vartheta}=0.88$ respectively.
Thus, the terminal speed depends strongly on shock location and
decreases with increasing shock location.

In Fig. (7b), we have plotted ${\vartheta}$ with ${\ell}$. 
The solid curves correspond to ${\dot m}_K=1.4$
and the dashed curves correspond to ${\dot m}_K=2.4$.
The shock locations $x_s$, are marked on each pair of
solid and dashed curves ($x_s=10r_g, 20r_g, 30r_g)$ in the Figure.
We see that in general ${\vartheta}$ increases with ${\ell}$.
We also see that, for particular values of $x_s$, ${\dot m}_K$
has very limited influence on ${\vartheta}$ (see each pair of solid and dashed
curves marked with values of $x_s=10, 20, 30$), and also
that with increasing $x_s$, ${\dot m}_K$ is less and less effective
in determining ${\vartheta}$. In particular, the solid and the
dashed curves
marked $x_s=10$ are distinguishable, but are increasingly
less distinguishable for $x_s=20$ and $x_s=30$. With the increase in $x_s$,
the inner edge of the Keplerian disc is increased. We also
know that the magnitude 
of Keplerian disc intensity is more, closer to the black
hole, \ie if
$x_s$ increases then the size of the Keplerian disc as well
as Keplerian luminosity
decreases. Thus with the increase of $x_s$, Keplerian
radiation will be less effective in determining ${\vartheta}$.
Similar to Fig. (5b), we also notice that if ${\ell}>0.11$ for the
pair of curves (solid and dashed) marked
$10$ (\ie $x_s=10$), then ${\vartheta}$ increases with decreasing
${\dot m}_K$,
but decreases with decreasing ${\dot m}_K$, for ${\ell}<0.11$.
Though this crossing over value of ${\ell}$, increases with increasing $x_s$,
\ie for $x_s=20r_g$, the crossing over occurs at ${\ell}=0.15$, and
for $x_s=30r_g$ this occurs at ${\ell}=0.24$.
We also see that, generally, ${\vartheta}$ decreases with increasing  $x_s$.
In Fig. (7c), ${\vartheta}$ is plotted with $x_s$, 
for ${\dot m}_K=0.5$. Various curves correspond to ${\ell}=0.24$ (solid),
${\ell}=0.18$ (dashed), ${\ell}=0.12$ (long-dashed) and ${\ell}=0.06$
(dashed-dotted). It is clear that ${\vartheta}$ decreases with $x_s$,
though for a fixed value of $x_s$, ${\vartheta}$ increases with ${\ell}$.
In Fig. (7d), ${\vartheta}$ is plotted with $x_s$, for constant values of
${\ell}=0.18$. Different curves corresponds to ${\dot m}_K=1.5$ (solid),
${\dot m}_K=3$ (dashed) and ${\dot m}_K=4.5$ (long-dashed).
The first thing to notice is the weak dependence of ${\vartheta}$ on 
${\dot m}_K$. We also notice that, for $x_s<22r_g$, ${\vartheta}$
decreases with increasing ${\dot m}_K$ and for $x_s>22r_g$,
${\vartheta}$ increases with increasing ${\dot m}_K$.
The reason for this is that the CENBOL intensity falls with increasing $x_s$. 
Thus for a larger $x_s$ the radiative moments due to the CENBOL at infinite 
distances are comparable to that due to the Keplerian disc. 
Hence with increasing ${\dot m}_K$, ${\vartheta}$
increases, exactly for the same reason as has been presented
while discussing Fig. (5b).
In general, one can conclude from Figs. (7a-7d), that
jets can be accelerated
to, up and around $90\%$ the velocity of light, provided
the $x_s{\rightarrow}
10r_g$---$20r_g$, and ${\ell}{\rightarrow}0.1$---$0.2$. As we discussed
earlier, such values of $\ell$ are not unreasonable when amplification of 
photon energy takes place due to Comptonization and $\tau \sim 1$. For smaller
$\tau$, amplification factor is higher (so that $\ell \sim 1$ is achievable)
but in that case, both the number of very energy photons goes down and
also the efficiency to deposite radiative momentum goes down dramatically.

\begin {figure}
\vbox{
\vskip -0.0cm
\hskip 0.0cm
\centerline{
\psfig{figure=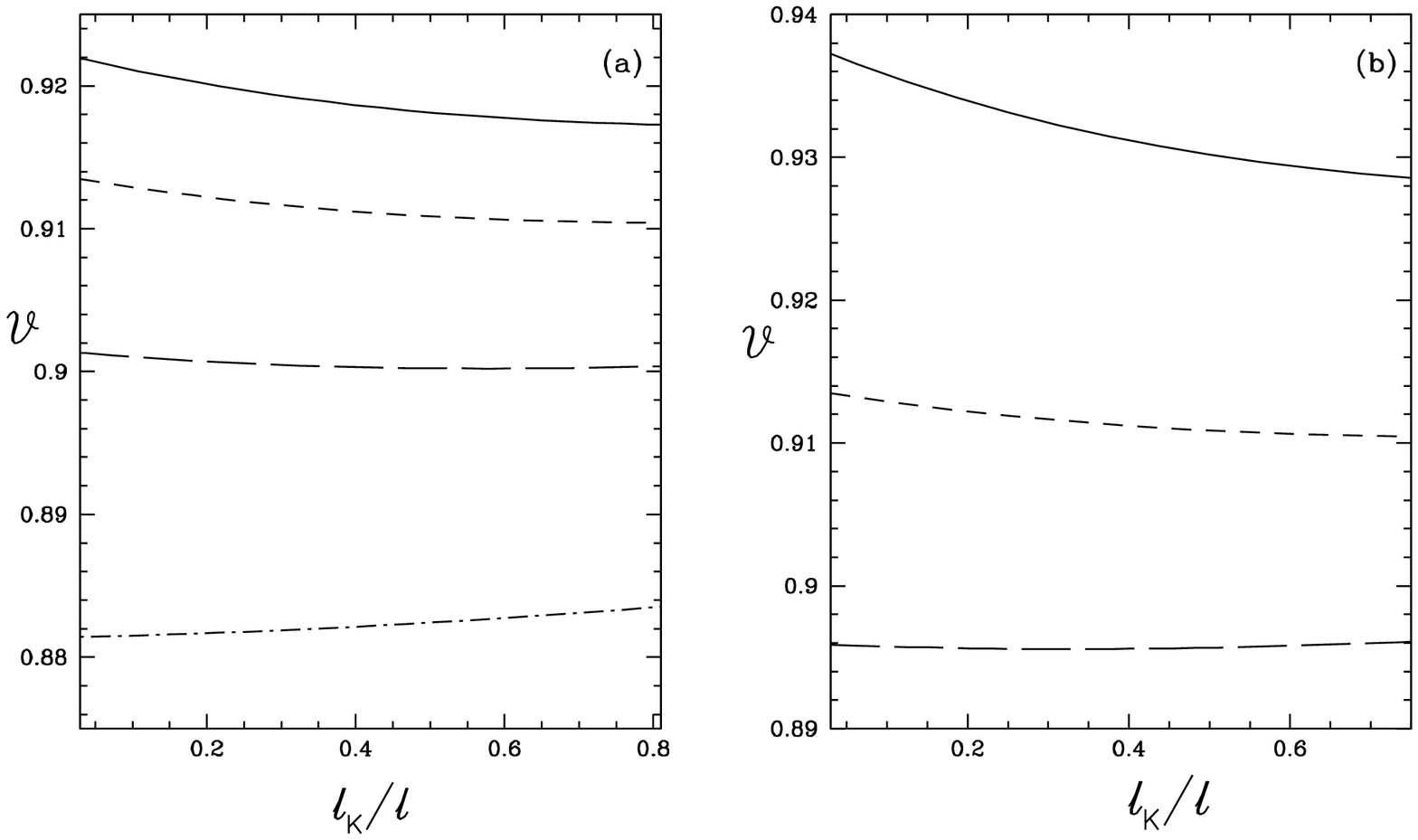,height=16truecm,width=15truecm}}}
\end{figure}
\begin{figure}
\vspace{-6.0cm}
\caption[] {(a) Variation of ${\vartheta}$ with ${\ell}_K/{\ell}$ for
$x_s=20$ for all the curves.
Various curves correspond to ${\ell}=0.3$ (solid),
${\ell}=0.24$ (dashed), ${\ell}=0.12$ (long-dashed)
and ${\ell}=0.12$ (dashed-dotted).
(b) Variation of ${\vartheta}$ with ${\ell}_K/{\ell}$
for ${\ell}=0.24$. 
Various curves corresponds to $x_s=10r_g$ (solid), $x_s=20r_g$
(dashed), $x_s=30r_g$ (long-dashed). Injection parameters $z_{in}=5r_g$,
$v_{in}=10^{-3}$.}
\end{figure}

In the preceding paragraphs of this subsection, we have studied
the issue of radiative acceleration of jets, and
on its dependence on three 
parameters namely $x_s$, ${\ell}$ and ${\dot m}_K$.
In case of CENBOL radiation, the information we have provided
is through its total luminosity,
but Keplerian luminosity is governed by two parameters
$x_s$ and ${\dot m}_K$ [see, Eq. (11)].
To have a better understanding, now we study
how the relative proportions of CENBOL and Keplerian luminosity affect
the terminal speed of jet.

In Fig. (8a), ${\vartheta}$ is plotted with ${\ell}_K/{\ell}$ ---
the ratio of Keplerian and CENBOL luminosities, for $x_s=20r_g$.
Various curves correspond to ${\ell}=0.3$ (solid), ${\ell}=0.24$ (dashed),
${\ell}=0.18$ (long-dashed) and ${\ell}=0.12$ (dashed-dotted).
The ratio of luminosities is kept less than one to mimic the hard state
of the accretion disc. 
We see that for ${\ell}=0.12$, ${\vartheta}$ increases with the increase
of Keplerian luminosity.
For higher CENBOL luminosities (see,  solid, dashed and long-dashed curves)
we observe that ${\vartheta}$ decreases with increasing Keplerian luminosity.
This has been addressed while discussing Fig. (5b), Fig. (6) and
Fig. (7b), \ie
for $x_s=20r_g$, if ${\ell}<0.15$, ${\vartheta}$ increases with the increase
of Keplerian luminosity.
It is thus clear that if $x_s=20r_g$ then
for ${\ell}$ higher than $0.15$, jets can be accelerated to very high terminal
velocities.
In order to study the dependence of shock location and the
Keplerian luminosity,
in Fig. (8b), we have plotted ${\vartheta}$ with ${\ell}_K/{\ell}$ for 
${\ell}=0.24$.
The Keplerian luminosity thus is increased from $3\%$
to $75\%$ of CENBOL luminosity. Various curves correspond to $x_s=10r_g$
(solid), $x_s=20r_g$ (dashed) and $x_s=30r_g$ (long-dashed).
For $x_s=10r_g$, ${\vartheta}$ is around $0.93$ but decreases with
the increase of Keplerian luminosity.
With the increase in shock location the terminal velocity is less
(${\vartheta}=0.91$ for $x_s=20r_g$), but at the same time, 
decrement of the terminal velocity due to the increase
of Keplerian luminosity is also less.
For $x_s=30r_g$, ${\vartheta}{\sim}0.895$ is lesser, but the
change due to the increase of Keplerian luminosity is even less
and ${\vartheta}$ remains almost constant.
Thus we conclude that if the shock location is between
$10r_g$---$20r_g$
then jets can be accelerated to terminal speeds above $90\%$
of the velocity of light, for disc luminosities around $20\%$
of the Eddington luminosity.

\section{Discussion and Concluding Remarks}

It is well known that high energy photons
can produce particle-antiparticle pairs close to the 
inner edge of a disk. If the photon energy $h{\nu} \gsim 2mc^2$, then an
electron-
positron pair may be created, where $h$ is the Planck's constant,
${\nu}$ is photon frequency and $m$ is the electron (or positron)
mass. If, on the other hand, electron and positron collide
it will annihilate each other to produce two Gamma-ray photons,
a process called pair annihilation. Clearly, to produce
electron-positron jets the pair-production process has to dominate
over pair annihilation. Workers (e.g. Mishra and Melia, 1993;
Yamasaki, Takahara and Kusunose, 1999)
discussed the production of electron-positron pairs from inner part
of the accretion disc.
Observationally, there are reports of pair dominated jets 
(Sunyaev \etal 1992; Mirabel and
Rodriguez 1998; Wardle \etal 1998) from galactic black hole candidates
to quasars. 
Though there is little doubt on the existence of pair dominated jets,
radiative acceleration of such jets on the other hand is a different
issue altogether.  If the number density of pairs created around the
the black hole is too high, then radiative acceleration would be ineffective.
In this present paper, we have ignored the details of
formation of the pair plasma jets and have only concentrated
on the radiative acceleration of optically thin pair dominated jets.

In this paper, we supplied the CENBOL intensity ($I_{C}$),
and the shock location $x_s$ as free, but reasonable parameters.
We have separately treated two velocity
variables, (i) the equilibrium velocity $v_{eq}$ and (ii)
the actual velocity $v$. While $v_{eq}$ decides how much velocity
is allowed before deceleration sets in, $v$ gives us what `net' value of 
velocity is achieved by actual acceleration. 
We have shown that if only CENBOL radiation dominates over the radiation
from the Keplerian disc, 
$v_{eq}{\sim}1$  is achieved within about few tens of Schwarzschild radii, 
but the same condition is achieved at over a thousand Schwarzschild radii for
much higher values of Keplerian luminosity. 
From Figs. (4a-4b), we have seen as the Keplerian accretion rate
is increased, $v_{eq}$ decreases in the range
$z{\rightarrow}15r_g$---$1500r_g$.

From Fig. (3d), we have seen that $({\tilde E}_K+{\tilde P}_K)>
2{\tilde F}_K$ in this very range, so in this range the radiation drag is
higher due to Keplerian radiation.
This means that for higher values of ${\ell}$ and for fixed $x_s$ \ie for
higher CENBOL intensity, the flow will tend to achieve very
high velocities within few tens of Schwarzschild radii, but at the same time
if Keplerian luminosity or ${\ell}_K$ is increased, then within the
range $15r_g$---$1500r_g$, Keplerian radiation will tend to reduce the high
velocities because of increased radiation drag.

From this study we generally conclude that, \\
(i) radiative acceleration of electron-positron jets
does achieve relativistic terminal speed. \\
(ii) The space dependant part of the radiative moments from 
the post-shock region, dominates the corresponding moments from the
Keplerian disc. \\
(iii) In general, the terminal speed of jets increases with increasing
post-shock luminosity. 
(iv) Post-shock radiative intensity decreases with
increasing shock location, so the terminal speed also decreases with
increasing shock location. \\
(v) Keplerian radiation has marginal effect in
determining the terminal speed. \\
(vi) Our calculations show that, if the shock in accretion is located at around
$10-20$ Schwarzschild radii, and if the post-shock luminosity is
about $10\%$ to $20\%$ of the Eddington luminosity, then
electron-positron jets can be accelerated to terminal speeds
above $90\%$ the speed of light.

This is supported in part by a grant from 
ESA Prodex project; ESA Contract no. 14815/00/NL/SFe(IC) [IC] and Department of Science and Technology, Govt. of India, through a grant no. SP/S2/K-15/2001
[SKC and SD].

{}

\begin{thebibliography}{}
\def\ref#1\par{\parshape=2 0in 14.5cm 1cm 13.5cm {#1} \par}
\parskip=0pt
\parindent=0pt

\bibitem[]{} 
Abramowicz, M. A., Piran, T., 1980, ApJ 241, L7

\bibitem[]{}
Bridle, A. H., Perley, R. A., 1984, ARA\&A 22, 319

\bibitem[]{}
Biretta, J. A., 1993, in Astrophysical Jets,
Burgerella, D., Livio, M., O\'Dea, C., eds, Volume 6 Space Telescope Science
Symposium series, Cambridge University Press, Cambridge, p. 263

\bibitem[]{}
Chakrabarti, S. K. 1989, ApJ 347, 365 (C89)

\bibitem[]{}
Chakrabarti, S. K., 1990, MNRAS 243, 610 (C90)

\bibitem[]{}
Chakrabarti, S.K. and Mandal, S. 2003, in New Views on MICROQUASARS, P. Durouchoux, Y. Fuchs and J. Rodriguez (Eds.), 117
(CSP:Kolkata)

\bibitem[]{}
Chakrabarti, S. K.,  Titarchuk, L., 1995, ApJ 455, 623 (CT95)

\bibitem[]{}
Chakrabarti, S. K., 1996, ApJ 464, 664 (C96)

\bibitem[]{}
Chakrabarti, S. K., Titarchuk, L., Kazanus, L., Ebisawa, K., 1996
A \& AS 120, 163 (CTKE96)

\bibitem[]{}
Chakrabarti, S. K., 1997, ApJ 484, 313 (C97)

\bibitem[]{}
Chakrabarti, S. K., 1998, in Chakrabarti, S. K., ed.,
Proc. Observational Evidence For 
Black Holes In The Universe. Kluwer Academic 
Publishers, Dordrecht, p. 19

\bibitem[]{}
Chakrabarti, S. K., 1999, A\&A 351, 185

\bibitem[]{}
Chattopadhyay, I., Chakrabarti, S. K., 2000, Int. Journ. Mod. Phys. D 9(1), 57

\bibitem[]{}
Chattopadhyay, I., Chakrabarti, S. K., 2002a, MNRAS 333, 454

\bibitem[]{}
Chattopadhyay, I., Chakrabarti, S. K., 2002b, in Durouchoux, P.,
Fuchs, Y., Rodriguez, J.,
eds, Proc. of the 4th Microquasar Workshop, CSP, Kolkata, p. 118

\bibitem[]{}
Chattopadhyay, I., Das, S., Mandal, S., Chakrabarti,
S. K., 2003, in Chakrabarti, S. K., Das, S., Basu, B.,Khan, M., eds,
Proc. Recent Trends in Astro and Plasma Physics in India,
CSP, Kolkata, p. 76 

\bibitem[]{} 
Das, T. K., Chakrabarti, S. K., 1999, Class. Quant. Grav. 16, 3879

\bibitem[]{}
Das, S., Chattopadhyay, I., Nandi, A., Chakrabarti, S. K., 2001, A\&A 379, 683 

\bibitem[]{}
Ebisawa, K., Titarchuk, L., Chakrabarti, S. K., 1996, PASJ 48, 59

\bibitem[]{}
Fukue, J., 1987, PASJ 39, 309

\bibitem[]{}
Fukue, J., 1996, PASJ 48, 631

\bibitem[]{}
Fukue, J., Tojyo, M., Hirai, Y., 2001, PASJ 53 555

\bibitem[]{}
Fukue, J., 2003, PASJ 55, 451

\bibitem[]{}
Gallo, E., Fender, R. P., Pooley, G. G., 2003, MNRAS, 344, 60

\bibitem[]{}
Icke, V., 1980, AJ 85(3), 329

\bibitem[]{}
Icke, V., 1989, A\&A 216, 294

\bibitem[]{}
Junor, W., Biretta, J. A., Livio, M., 1999, Nature, 401, 891

\bibitem[]{}
Mihalas, D., Mihalas, B. W., 1984, Foundations of Radiation
Hydrodynamics, Oxford University Press, Oxford (MM84)

\bibitem[]{}
Kato, S., Fukue, J., Mineshige, S., 1998, Black-hole accretion disks,
Kyoto University Press, Kyoto (K98)

\bibitem[]{}
Liang, E. P. T., Thompson, K. A., 1980, ApJ 240, 271L

\bibitem[]{}
Lynden-Bell, D., 1978, Physica Scripta 17, 185

\bibitem[]{}
Mirabel, I. F., Rodriguez, L. F., 1994, Nat 371, 46

\bibitem[]{}
Mirabel, I. F., Rodriguez, L. F., 1998, Nat 392, 673

\bibitem[]{}
Mishra, R., Melia, F., 1993, ApJ, 419 L25

\bibitem[]{}
Novikov, I. D., Thorne, K. S., 1973, in Black Holes,
C. Dewitt and B. Dewitt (eds.),  
Gordon and Breach, New York, p. 343 (NT73)

\bibitem[]{}
Paczy\'nski, B., Wiita, P., 1980, A\&A 88, 23

\bibitem[]{}
Piran, T., 1982, ApJ 257, L23

\bibitem[]{}
Shakura, N. I., Sunyaev, R. A., 1973, A\& A 24, 337

\bibitem[]{}
Sikora, M., Wilson, D. B., 1981, MNRAS 197, 529

\bibitem[]{}
Smith, D. M., Heindl, W. A., Markwardt, C. B., Swank, J. H., 2001, ApJ 554, L41

\bibitem[]{}
Smith, D. M., Heindl, W. A., Swank, J. H., 2002, ApJ 569, 362

\bibitem[]{}
Sol, H., Pelletier, G., Ass{\^e}o, E., 1989, MNRAS 237, 411

\bibitem[]{}
Sunyaev, R., Churazov, E., Gilfanov, M., Dyachkov, A.,
Khavenson, N., Grebenev, S., Kremnev, R., Sukhanov, K., 1992, ApJ,
389, L75

\bibitem[]{} Yamasaki, T., Takahara, F., Kusunose, M., 1999, ApJ, 523, L21

\bibitem[]{}
Wardle, J. F. C., Homan, D. C., Ojha, R.,  Roberts, D. H., 1998, Nat
395, 457

\bibitem[]{}
Zensus, J. A., Cohen, M. H., Unwin S. C., 1995, ApJ 443, 35

\end{thebibliography}
\end{document}